\documentclass[12pt]{iopart}
\usepackage{iopams}  
\usepackage[dvips,dvipdfm]{graphicx}

\newcommand{\mR}{\mathbb{R}}

\newcommand{\out}{y}
\newcommand{\weight}{w}
\newcommand{\ltsim}{\protect\raisebox{-0.5ex}{$\:\stackrel{\textstyle <}
	{\sim}\:$}}

\newcommand{\img}{{\rm i}}

\begin{document}

\title
{Learning of correlated patterns by simple perceptrons}

\author{Takashi Shinzato
\ and Yoshiyuki Kabashima
}

\address{
Department of Computational Intelligence and Systems Science\\
Tokyo Institute of Technology, Yokohama 226-8502, Japan\\
shinzato@sp.dis.titech.ac.jp, kaba@dis.titech.ac.jp}

\begin{abstract}
Learning behavior of simple perceptrons is analyzed for a teacher-student scenario in which output labels are provided by a teacher network  for a set of possibly correlated input patterns, and such that teacher and student networks are of the same type. Our main concern is the effect of statistical correlations among the input patterns on  learning performance. For this purpose, we extend to the teacher-student scenario a methodology for analyzing randomly labeled patterns recently developed in {\em J. Phys. A: Math. Theor.} {\bf 41}, 324013 (2008). This methodology is used for analyzing situations in which orthogonality of the input patterns is enhanced in order to optimize the learning performance.  
\end{abstract}



\maketitle

\section{Introduction}
Learning from examples is a fundamental technique  
for analyzing real-world data, 
and simple perceptrons are included in widely used devices for 
this purpose. 
In the last two decades, the structural similarity 
between statistical learning and statistical mechanics of 
disordered systems has been commonly recognized \cite{Levin1990}. 
This similarity has promoted statistical mechanical analysis 
of perceptron learning \cite{Watkin1993,Engel2001,Nishimori2001}. 
Such research has led to the discovery 
of various learning behaviors of perceptrons
and the development of computationally feasible 
approximate algorithms that had not previously been 
known in conventional learning research 
\cite{Gardner1988,Gyorgyi1990PRA,Gyorgyi1990,Krauth1989,
Krauth1989JPA,Opper1996,Kabashima2003,Uda2005,Braunstein2006}.

Numerous studies have been published on perceptron 
learning. 
However, there still remain several research directions to explore. 
Learning from correlated patterns is a typical 
example.  As a first step in
this direction, the authors recently developed methodologies 
to analyze learning from randomly labeled patterns that are 
correlated in a certain manner on the basis of a formula 
involving rectangular random matrices \cite{Kabashima2007,
Shinzato2008}. 
This paper is concerned with a second step; more precisely, 
we extend the methodologies developed for
randomly labeled patterns to cases of a 
teacher-student scenario in which output labels
are provided by a teacher network, and both teacher and student  networks are of the same  type.

In earlier studies, asymptotic behavior of learning curves and a critical 
pattern ratio of perfect learning in which the teacher network is 
completely identified from a reference data set of the same 
order as the network size, have been assessed for continuous and discrete weights, 
respectively, for the case of independently and identically distributed 
(i.i.d.) patterns \cite{Gyorgyi1990PRA,Gyorgyi1990}. 
Therefore, our main concern herein is how these assessments
are influenced by correlations among input patterns. 
Recent deeper understanding of the relations among learning, communication and 
information theories has suggested that a perceptron can be a
useful building block for various coding 
schemes \cite{Hosaka2002,Kinzel2002,Mimura2006}. 
The analysis herein may also be a useful guideline for developing 
efficient schemes to be used in information and communication engineering.

This paper is organized as follows.
In the next section, we define the model that 
we shall investigate. 
In section \ref{sec3}, the main section of 
this manuscript, we will extend a scheme to 
handle correlated patterns, utilizing a formula involving 
rectangular random matrices, which was developed in 
\cite{Kabashima2007,Shinzato2008}, 
to perceptron learning of a teacher-student scenario. 
In section \ref{sec4}, the extended scheme will be applied to 
several examples. 
The final section is devoted to a summary and future work.

\section{Model definition \label{sec2}}
For an $N$ dimensional input pattern vector 
$\vec{x}$, a single layer perceptron of weight 
$\vec{\weight}$ of dimension $N$ returns a binary 
output $\out \in \{+1,-1\}$ given by
\begin{eqnarray}
\out={\rm sgn}\left(\frac{1}{\sqrt{N}}\vec{x}^{\rm T}
\vec{\weight} \right)=\left\{
\begin{array}{ll}
1,&N^{-1/2}\vec{x}^{\rm T}\vec{\weight} >0, \\
-1,&{\rm otherwise,}
\end{array}
\right. 
\label{eq1}
\end{eqnarray}
where ${\rm T}$ denotes the matrix transpose and the
prefactor $1/\sqrt{N}$ is introduced 
to keep relevant variables  $O(1)$ as $N \to \infty$. 
Let us suppose a situation in which a student perceptron infers 
the weight vector of a teacher perceptron, $\vec{\weight}_0$, 
based on a given reference data set 
$\xi^p=\left\{\left(\vec{x}_1^{\rm T},\out
_1\right),\left(\vec{x}_2^{\rm T},\out_2\right)\ldots, 
\left(\vec{x}_p^{\rm T},\out_p\right)\right\}$, where
output labels are provided by the teacher
as $\out_\mu={\rm sgn}\left(N^{-1/2}\vec{x}_\mu^{\rm T}
\vec{\weight}_0 \right)$ $(\mu=1,2,\ldots,p)$. 
The problem we consider here is how 
the inference accuracy depends on the pattern ratio
$\alpha=p/N$ and correlations in the pattern matrix 
$X=N^{-1/2}(\vec{x}_1,\vec{x}_2,\ldots,\vec{x}_p)^{\rm T}$ as 
$N$ and $p$ tend to infinity, while keeping $\alpha=p/N$ finite.

As the basis for our analysis, 
we introduce a representation of the singular value decomposition 
\begin{eqnarray}
X&=&UDV^{\rm T},
\label{SVD}
\end{eqnarray}
where $D={\rm diag}(d_k)$ is a $p\times N$ diagonal matrix 
consisting of singular values $d_k$ $(k=1,2,\ldots,\min(p,N))$, 
and $U$, $V$ denote $p\times p$ and $N\times N$ orthogonal matrices, 
respectively. Linear algebra guarantees that any $p \times N$ matrices
can be decomposed according to equation (\ref{SVD}). 
The singular values are linked to the eigenvalues $\lambda_k$
$(k=1,2,\ldots,N)$ of 
the correlation matrix $X^{\rm T} X$ via
$\lambda_k=d_k^2$ $(k=1,2,\ldots,{\rm min}(p,N))$ and $0$ otherwise, 
where ${\rm min}(p,N)$ denotes the lesser value of $p$ and $N$. 
Orthogonal matrices $U$ and $V$ constitute the right and left 
eigen-bases of $X$, respectively; i.e., 
they are the eigen-bases of $XX^{\rm T}$ and $X^{\rm T}X$. 

To handle the correlations in $X$ somewhat analytically, we assume hereinafter that 
the following two properties hold for
the pattern matrix $X$: 
\begin{enumerate}
\item The eigenvalue spectrum of the correlation matrix $X^{\rm T}X$, 
$\rho_{X^{\rm T}X}(\lambda)=N^{-1}\sum_{k=1}^N \delta(\lambda-\lambda_k)$, 
tends to a certain specific distribution 
$\rho(\lambda)$ in the limit as $N \to \infty$ for typical samples
of $X$. Controlling $\rho(\lambda)$ allows us to characterize various 
second-order correlations in $X$. 
\item $U$ and $V$ are independently generated from 
the uniform distributions of $p \times p$ and 
$N \times N$ orthogonal matrices (the Haar measures), 
respectively. This assumption makes it possible to 
characterize the correlations in $X$ using only 
the eigenvalue spectrum $\rho(\lambda)$. 
\end{enumerate}

\section{Analytical scheme \label{sec3}}
\subsection{Expression for the average free energy}
Given $\xi^p$, the volume of weight vectors that are compatible 
with $\xi^p$, which serves as the partition function of 
the current system, can be expressed as
\begin{eqnarray}
Z\left(\xi^p\right)&=&\sum_{\vec{\weight}}
P\left(\vec{\weight}\right)\prod_{\mu=1}^p
\Theta\left(\frac{\out_\mu}{\sqrt{N}}\sum_{k=1}^Nx_{\mu k}\weight_{k}\right), 
\label{gardner}
\end{eqnarray}
where $P(\vec{\weight})$ represents 
the prior distribution of $\vec{\weight}$
and $\Theta(x)=1$ for $x \ge 0$ and $0$, otherwise. 
The conventional scheme of statistical mechanics of 
disordered systems indicates that 
typical properties of learning can be examined by 
assessing the average free energy 
\begin{eqnarray}
\Phi =-\frac{1}{N} \left [ \log Z\left (\xi^p  \right ) 
\right ]_{\xi^p}. 
\label{entropy}
\end{eqnarray}
Here, $\left[ \cdots \right]_{\xi^p}$ represents 
an average taken over the reference data set $\xi^p=\{X,\vec{\out}\}$ 
with respect to a distribution 
\begin{eqnarray}
P(\xi^p)=P(X) \sum_{\vec{\weight}_0} P(\vec{\weight}_0)
\prod_{\mu=1}^p
\Theta\left(\frac{\out_\mu}{\sqrt{N}}\sum_{k=1}^Nx_{\mu k}\weight_{0k}\right)
=P(X)Z(\xi^p), 
\label{Pxi}
\end{eqnarray}
where $P(X)$ denotes the distribution of the pattern matrix $X=UDV^{\rm T}$. 
This implies that equation (\ref{entropy}) can be evaluated as
\begin{eqnarray}
\Phi =-\lim_{n \to 1} \frac{\partial }{\partial n} \frac{1}{N} 
\log \left (\sum_{\xi^p}P(X) Z^n(\xi^p) \right ). 
\label{n1replica}
\end{eqnarray}

\subsection{Replica analysis}
Equation (\ref{n1replica}) can be evaluated by the replica method. 
For this, we first evaluate 
$\overline{Z^n(\xi^p) }=\sum_{\xi^p}P(X)Z^n(\xi^p)$
for $n=1,2,\ldots$ utilizing the expression
\begin{eqnarray}
Z(\xi^p)&=&\sum_{\vec{\weight}} 
\int \prod_{\mu=1}^p d \Delta_\mu 
\Theta (\out_\mu \Delta_\mu)
\delta\left (\Delta_\mu-\frac{1}{\sqrt{N}}
\sum_{k=1}^N x_{\mu k}\weight_k \right ) \cr
&=&\sum_{\vec{\weight}}\int d \vec{u}
\prod_{k=1}^N P(\weight_k)
\times  \prod_{\mu=1}^p \widetilde{\Theta}(\out_\mu u_\mu)
\times e^{-\img \vec{u}^T X \vec{\weight} }, 
\label{auxexpressionV}
\end{eqnarray}
where $\img=\sqrt{-1}$, 
$\widetilde{\Theta}(x)=(2 \pi)^{-1}\int dt \Theta(t)e^{\img tx}$
and $\vec{u}=(u_1,u_2,\ldots,u_p)^{\rm T}$. 
We have assumed a factorizable prior $P(\vec{\weight})=\prod_{k=1}^N 
P(\weight_k)$ for analytical tractability. Taking $n$-th powers, for 
$n(=1,2,\ldots)$, equation (\ref{auxexpressionV})
yields an expression 
\begin{eqnarray}
\exp \left [-\img \sum_{a=1}^n \vec{u}_a^{\rm T} X \vec{\weight}_a \right ]
=\exp \left [-\img \sum_{a=1}^n (U^{\rm T}\vec{u}_a)^{\rm T} D
(V^{\rm T}\vec{\weight}_a )\right ]. 
\label{UDV_replica}
\end{eqnarray}
For evaluating the average of this equation with respect to $X$, 
it is useful to note that for fixed sets of 
dynamical variables $\{\vec{u}_a\}=\{\vec{u}_1,\vec{u}_2,\ldots,\vec{u}_n\}$
and $\{\vec{\weight}_a\}=\{\vec{\weight}_1,\vec{\weight}_2,\ldots,
\vec{\weight}_n\}$, 
$\widetilde{\vec{u}}_a=U^{\rm T} \vec{u}_a$ and 
$\widetilde{\vec{\weight}}_a=V^{\rm T} \vec{\weight}_a$ 
behave as continuous random variables which satisfy the strict 
constraints
\begin{eqnarray}
\frac{1}{N} \widetilde{\vec{\weight}}_a^{\rm T} \widetilde{\vec{\weight}}_b
&=&\frac{1}{N} {\vec{\weight}}_a^{\rm T} {\vec{\weight}}_b = q_{ab}^w, \\
\frac{1}{p} \widetilde{\vec{u}}_a^{\rm T} \widetilde{\vec{u}}_b
&=&\frac{1}{p} {\vec{u}}_a^{\rm T} {\vec{u}}_b = q_{ab}^u, 
\end{eqnarray}
$(a,b=1,2,\ldots,n)$ when $U$ and $V$ are independently sampled 
from the Haar measures. 
This indicates that $\overline{Z^n(\xi^p)}$ can be evaluated 
by the saddle point method with respect to 
the macroscopic order parameters ${\cal Q}^w=(q_{ab}^w)$ and 
${\cal Q}^u=(q_{ab}^u)$ in the limit as 
$N,p \to \infty$,  keeping $\alpha=p/N \sim O(1)$. 
Furthermore, due to the intrinsic permutation symmetry with 
respect to the replica indices $a=1,2,\ldots,n$, 
it is natural to assume that the relevant saddle point 
is replica symmetric (RS).  This assumption can be expressed as 
\begin{eqnarray}
q_{ab}^w=\left \{
\begin{array}{ll}
\chi_\weight+q_\weight, & (a=b), \cr
q_\weight, & (a\ne b), 
\end{array}
\right .
\quad 
q_{ab}^u=\left \{
\begin{array}{ll}
\chi_u-q_u, & (a=b), \cr
-q_u, & (a\ne b), 
\end{array}
\right .
\label{RS}
\end{eqnarray}
which yields 
\begin{eqnarray}
&&\frac{1}{N} \log \left (
\int {\cal D}XP(X)e^{-\img \sum_{a=1}^n \vec{u}_a^{\rm T}
X \vec{\weight}_a } \right )\cr
&&=(n-1)F(\chi_\weight,\chi_u)+F(\chi_\weight+nq_\weight,\chi_u-nq_u), 
\label{RS_avX}
\end{eqnarray}
for fixed sets of $\{\vec{\weight}_a\}$ and $\{\vec{u}_a\}$, 
where 
\begin{eqnarray}
F(x,y)&=&\mathop{\rm Extr}_{\Lambda_x,\Lambda_y}
\left \{
-\frac{\alpha-1}{2}\log \Lambda_y
-\frac{1}{2}\left \langle \log (\Lambda_x \Lambda_y+\lambda) \right \rangle 
+\frac{\Lambda_x x}{2}+\frac{\alpha \Lambda_yy}{2}
\right \}
\cr
&& -\frac{1}{2}\log x-\frac{\alpha}{2}\log y -\frac{\alpha+1}{2}, 
\label{F_function}
\end{eqnarray}
and $\int {\cal D}X$ denotes integration with respect to 
the pattern matrix $X$ \cite{Kabashima2007,Shinzato2008,Kabashima2008}. 
$\left \langle (\cdots) \right \rangle$ means 
an average of $(\cdots)$ with respect to 
$\rho(\lambda)$. 
$\mathop{\rm Extr}_{x}\{\cdots\}$ denotes the operation of 
extremization with respect to $x$, which 
corresponds to the saddle point evaluation of a certain 
complex integral and does not refer to 
maximization or minimization. 

Equation (\ref{RS}) and assessment of the volumes of the dynamical 
variables yield a saddle point evaluation of $\overline{Z^n(\xi^p)}$
for $n=1,2,\ldots$. However, the functional form 
of $\overline{Z^n(\xi^p)}$ that we obtain can be defined for real 
values of $n$ as well. Therefore, we analytically 
continue the expression from $n =1,2,\ldots$
to $n \in \mR$ to evaluate equation (\ref{n1replica}). 
For $n \to 1$, the normalization constraint $\overline{Z(\xi^p)}=
\sum_{\xi^p} P(X) Z(\xi^p)=\sum_{\xi^p}P(\xi^p)=1$
implies that relations 
\begin{eqnarray}
&&\chi_\weight+q_\weight=T_\weight=\sum_{\weight}P(\weight)
\weight^2, \quad \chi_u-q_u=0, \\
&& \frac{\partial F(\chi_\weight+q_\weight,\chi_u-q_u)}{\partial \chi_\weight}=0, \quad
\frac{\partial F(\chi_\weight+q_\weight,\chi_u-q_u)}{\partial \chi_u}=
\frac{1}{2}\left \langle \lambda \right \rangle T_w, 
\label{constraints}
\end{eqnarray}
must hold.
These yield a formula for calculating the average free energy as
\begin{eqnarray}
\Phi =-\mathop{\rm Extr}_{q_\weight,q_u}\left \{
{\cal A}_{\weight u}(q_\weight,q_u)+{\cal A}_\weight
(q_\weight)+\alpha {\cal A}_u(q_u)
\right \}, 
\label{replica_entropy}
\end{eqnarray}
where 
\begin{eqnarray}
{\cal A}_{w u}(q_\weight,q_u)=F(T_\weight-q_\weight,q_u)+
\frac{1}{2}\left \langle \lambda \right \rangle T_\weight q_u, 
\label{Awu}
\end{eqnarray}
\begin{eqnarray}
{\cal A}_{\weight }(q_\weight)=
\mathop{\rm Extr}_{\widehat{q}_\weight}
\left \{-\frac{\widehat{q}_\weight q_\weight}{2}+\int Dz 
{\cal P}(z;\widehat{q}_\weight)\log {\cal P}(z;\widehat{q}_\weight) 
\right \}, 
\label{Aw}
\end{eqnarray}
and 
\begin{eqnarray}
{\cal A}_{u}(q_u)=
\mathop{\rm Extr}_{\widehat{q}_u}
\left \{
-\frac{\widehat{q}_u q_u}{2}+ 2 \int Dz H(\gamma z)\log H(\gamma z)
\right \}, 
\label{Au}
\end{eqnarray}
given a particular eigenvalue spectrum $\rho(\lambda)$. 
Here, $Dz=dz e^{-z^2/2}/\sqrt{2 \pi}$ represents the Gaussian measure, 
$H(u)=\int_u^{+\infty}Dz$, 
${\cal P}(z;\widehat{q}_\weight)=\sum_{\weight}P(\weight) \exp \left [-
\widehat{q}_\weight \weight^2/2 +
\sqrt{\widehat{q}_\weight}z w \right ]$ and 
$\gamma=\sqrt{\widehat{q}_u/
\left (\left \langle \lambda \right \rangle T_\weight/\alpha
-\widehat{q}_u \right )}$. 
Equations (\ref{replica_entropy})--(\ref{Au}) contain the main results of this 
paper. 

Two points are noteworthy here. Firstly, $q_\weight$ being determined by the 
saddle point condition of equation (\ref{replica_entropy}) physically 
means that there is a typical overlap $N^{-1} \vec{\weight}_0^{\rm T} \vec{\weight}$
between the teacher $\vec{\weight}_0$ 
and student $\vec{\weight}$ perceptrons after learning. 
Therefore the learning performance can be assessed by solving the 
saddle point problem of equation (\ref{replica_entropy}). 
In addition, equation (\ref{replica_entropy}) itself 
represents the mutual information (per component) between 
$\vec{\weight}$ and $\vec{\out}$ for a given 
typical pattern matrix $X$, which 
measures the quantity of  information about $\vec{\weight}$ 
that can typically be gained from output labels $\vec{\out}$
when $X$ is fixed, and so is useful for characterizing potential 
capabilities of simple perceptrons when they are used 
for communication purposes. 
This is also linked to the typical entropy (per component) $S$ of 
the posterior distribution 
$P(\vec{\weight}|\xi^p)=Z^{-1}(\xi^p)P(\vec{\weight})
\prod_{\mu=1}^p
\Theta\left(\out_\mu N^{-1/2}\sum_{k=1}^Nx_{\mu k}\weight_{k}\right)$ via
\begin{eqnarray}
S=S_0-\Phi, 
\label{entropy_freeenergy}
\end{eqnarray}
where $S_0=-N^{-1}\sum_{\vec{\weight}}P(\vec{\weight})\log P(\vec{\weight})$ 
is the entropy (per component) of the prior 
distribution $P(\vec{\weight})$. 
Secondly, the assumption that the teacher and student networks
are of the same type corresponds to the Nishimori condition 
known in spin glass research, 
which implies that 
the RS solution constructed above 
is expected to be correct \cite{Nishimori1981,Nishimori1993,Nishimori2001b}. 
Therefore, we do not proceed to the replica symmetry breaking 
(RSB) analysis here. Treatment in a more general setting, including 
the local stability analysis of the RS solution and 
the expression of the 1RSB free energy, can be found 
in references \cite{Kabashima2007,Shinzato2008}.

\section{Examples\label{sec4}}
\subsection{Independently and identically distributed patterns}
In order to show consistency with existing results, 
we firstly apply the scheme that we have developed here to the case
of i.i.d. patterns, in which entries of 
the pattern matrix $X$ are independently generated from 
an identical distribution with mean zero and variance $1/N$. 
In the current framework, this case is 
characterized by the Mar$\breve{\rm c}$enko-Pastur distribution 
\begin{eqnarray}
\rho(\lambda)&=&[1-\alpha]^+\delta(\lambda)+
\frac{\sqrt{[\lambda-\lambda_-]^+[\lambda_+-\lambda]^+}}{2\pi\lambda}, 
\label{MP}
\end{eqnarray}
where $\lambda_{\pm}=\left(1\pm\sqrt{\alpha}\right)^2$ and 
\begin{eqnarray}
\left[x\right]^+=\left \{
\begin{array}{ll}
x, & (x \ge 0),  \cr
0, & (x < 0). 
\end{array}
\right . 
\end{eqnarray}
Plugging equation (\ref{MP}) into 
equation (\ref{F_function}) yields
\begin{eqnarray}
F(x,y)=-\frac{\alpha}{2} xy. 
\label{Fxy}
\end{eqnarray}
Applying this to equation (\ref{Awu}) and utilizing the relation 
$\left\langle\lambda\right\rangle=\alpha$
which holds for equation (\ref{MP})
yields the result that
${\cal A}_{wu}(q_\weight,q_u)=\alpha q_\weight q_u/2$. This implies that
$\widehat{q}_u=q_\weight$ at the extremum, where 
$\widehat{q}_u$ is the auxiliary variable in equation (\ref{Au}). 
These conditions mean that the average free energy can be calculated as 
\begin{eqnarray}
\Phi =&-&\mathop{\rm Extr}_{q_\weight,\tilde{q}_\weight}
\left\{\int Dz \log 
\left (\sum_\weight P(\weight) e^{-\frac{1}{2}\widehat{q}_\weight \weight^2 
+\sqrt{\widehat{q}_\weight}z \weight} \right )
-\frac{1}{2}\widehat{q}_\weight q_\weight 
\right . \cr
&+& \left . 2\alpha\int_{-\infty}^\infty 
Dz H\left(\sqrt{\frac{q_\weight}{T_\weight-q_\weight}}z \right)
\log H\left(\sqrt{\frac{q_\weight}{T_\weight-q_\weight}}z \right)
\right\}, 
\end{eqnarray}
which is equivalent to the known expression for the free energy 
of the teacher-student scenario with i.i.d. patterns \cite{Opper1996}.

\subsection{Asymptotic learning curve for spherical weights}
The relation between a measure of learning  performance 
and the amount of reference data $p$ or the pattern ratio $\alpha$ is 
sometimes termed a {\em learning curve}. 
In statistical learning theory, the asymptotic behavior 
of learning curves is frequently examined, which is, however, 
limited mostly to the cases of i.i.d. 
patterns \cite{Baum1990,Amari1992,Murata1994,Watanabe2001}. 
We here employ the methodology that has been developed for the analysis of 
the asymptotic learning curve in order to investigate 
the effect of correlations in the pattern matrix. 
Investigations of this kind may be useful for {\em active learning }
or {\em experimental design} contexts in which the pattern matrix can be 
designed to optimize learning performance 
\cite{Fukumizu1994,Sollich1994,Seeger2008}. 

As a representative example, let us consider the 
case of spherical weights $P(\vec{\weight})\propto 
\delta(|\vec{\weight}|^2-N)$, which implies that $T_\weight=1$. 
For generality, we investigate a model for which the 
second-order correlations of the 
pattern matrix $X$ are characterized by an eigenvalue spectrum
\begin{eqnarray}
\rho(\lambda)=(1-\kappa)\delta(\lambda)+\kappa
\widetilde{\rho}(\lambda), 
\label{general_eigen_value}
\end{eqnarray}
where $\widetilde{\rho}(\lambda)$ is a distribution, 
the support of which is defined over a certain 
region of $\lambda > 0$. 
$0\le \kappa \le 1$ is introduced to include the possibility 
of rank deficiency of the pattern matrix.

For this situation, we evaluate $q_\weight$, which 
serves as a performance measure representing the
overlap between teacher and student perceptrons, 
for $\alpha \gg 1$ solving the 
saddle point problem of equation (\ref{replica_entropy}). 
For spherical weights, $\Lambda_\weight$, which is the 
counterpart of $\Lambda_x$ in equation (\ref{F_function}) for
$x=T_\weight-q_\weight=1-q_\weight$, is always fixed to unity 
at the saddle point. This yields four coupled equations
relevant to the calculation of $q_\weight$ thus:
\begin{eqnarray}
\widehat{q}_u&=&\frac{\left \langle 
\lambda \right \rangle }{\alpha}+
\frac{2}{\alpha} \frac{\partial F(1-q_\weight,q_u)}{\partial q_u}
=\frac{\left \langle \lambda \right \rangle }{\alpha}+
\Lambda_u-\frac{1}{q_u}, \label{qhu}\\
q_u&=&\left (1-\frac{\kappa}{\alpha} \right )\frac{1}{\Lambda_u}
+\frac{\kappa}{\alpha}\left \langle \frac{1}{\Lambda_u+\lambda} 
\right \rangle_{\widetilde{\rho}}, \label{qu1} \\
q_u&=&\frac{\alpha }{\pi \left \langle \lambda \right \rangle 
\sqrt{1-\alpha \widehat{q}_u/\left \langle \lambda \right \rangle }}
\int Dz 
\frac{e^{-\alpha \widehat{q}z^2/(2\left \langle \lambda \right \rangle)}}
{H\left (\sqrt{\alpha \widehat{q}_u z/
\left \langle \lambda \right \rangle } \right )}, 
\label{qu2}\\
1-q_\weight&=&(1-\kappa)+\kappa
\left \langle \frac{\Lambda_u}{\Lambda_u+\lambda} 
\right \rangle_{\widetilde{\rho}}, \label{qw}
\end{eqnarray}
where $\left \langle \cdots \right \rangle_{\widetilde{\rho}}$
represents an average with respect to $\widetilde{\rho}(\lambda)$. 
For $\alpha \gg 1$, equations (\ref{qu1}) and (\ref{qu2})
yield asymptotic relations 
$q_u \simeq (1-\kappa/\alpha)/\Lambda_u$ and 
$1-\alpha \widehat{q}_u/
\left \langle \lambda \right \rangle
\simeq (c \alpha /\pi \left \langle \lambda \right \rangle)^2/q_u^{2}
\simeq (c \alpha /\pi \left \langle \lambda \right \rangle)^2
\Lambda_u^2$, where
$c=\int Dz e^{-z^2/2}/H(z) \simeq 2.263$. 
Inserting these relations into equation (\ref{qhu}) yields
$\Lambda_u \simeq \kappa 
\left \langle \lambda \right \rangle
(\pi/c)^2/\alpha^2 \simeq 1.926\kappa
 \left \langle \lambda \right \rangle/\alpha^2$. 
From this result and equation (\ref{qw}), we obtain the asymptotic learning 
curve  
\begin{eqnarray}
q_{\weight} \simeq \kappa -
\frac{1.926 \kappa^2 \left \langle \lambda \right \rangle
\left \langle \lambda^{-1}
\right \rangle_{\widetilde{\rho}} 
}{  \alpha^2} + O(\alpha^{-3}). 
\label{asymptotic}
\end{eqnarray}

Two issues are noteworthy here. 
Firstly, in the current model, $\kappa$ denotes 
a fraction of the relevant dimensions that the pattern
matrix $X$ spans. 
Convergence $q_\weight \to \kappa$ as $\alpha \to \infty$ in
equation (\ref{asymptotic}) indicates that 
weights concerning the relevant dimensions are
correctly identified, while no information is 
obtained from the irrelevant dimensions for perceptron 
learning. The rate of convergence scales 
as $O(\kappa^2 \alpha^{-2} )=O((\kappa N)^2/p^2)$, 
which indicates that the irrelevant dimensions do
not affect the learning performance of the relevant weights. 
This is in accordance with existing results 
for singular statistical models in which some of the
eigenvalues of the Fisher information 
matrix vanish, similar to cases of equation 
(\ref{general_eigen_value}) such that $0 \le \kappa < 1$ \cite{Watanabe2001}. 
Secondly, the inequality $\left \langle \lambda^{-1}
\right \rangle_{\widetilde{\rho}} \ge
\left \langle \lambda
\right \rangle_{\widetilde{\rho}}^{-1}=\kappa \left \langle \lambda
\right \rangle^{-1}$, 
which holds because $\lambda$ is positive and 
$\left \langle \lambda \right \rangle=
\kappa \left \langle \lambda 
\right \rangle_{\widetilde{\rho}}$ is satisfied
by equation (\ref{general_eigen_value}), 
implies that $q_{\weight}$ is asymptotically 
bounded above:
\begin{eqnarray}
q_{\weight } \ltsim \kappa -\frac{1.926 \kappa^3 }{ \alpha^2}
+ O(\alpha^{-3}), 
\label{upperbound}
\end{eqnarray}
where equality holds when $\left \langle \lambda^{-1}
\right \rangle_{\widetilde{\rho}}
=\left \langle \lambda 
\right \rangle_{\widetilde{\rho}}^{-1}$ is satisfied. 
This property is asymptotically satisfied for the i.i.d. patterns
since equation (\ref{MP}) yields 
$\left \langle \lambda^{-1} \right \rangle=(\alpha-1)^{-1}
\simeq \alpha^{-1}=\left \langle \lambda \right \rangle^{-1}$
for $\alpha \gg 1$. 
In addition, $\kappa=1$ holds for $\alpha \gg 1$ of 
the i.i.d. patterns, which maximizes
the value of convergence $\kappa$ to unity, 
reproducing the known asymptotic learning 
curve $\cos^{-1}(q_w)/\pi \simeq 0.625/\alpha$ \cite{Gyorgyi1990}. 
Therefore, the i.i.d. patterns are asymptotically optimal 
for the leading order of the learning curve 
although certain improvements can be gained for the next order
by optimally designing the pattern matrix.

\subsection{Presumed optimal performance in the non-asymptotic region}
The above argument characterizes the optimal learning performance
of simple perceptrons in the asymptotic region $\alpha \gg 1$. 
On the other hand, in information theory, 
it is known that when $\vec{\weight}$ is transformed to $\vec{\out}$ via
$\vec{\out}=X \vec{\weight}+\vec{n}$, 
where $\vec{n}$ is an noise vector whose components are
i.i.d. Gaussian random numbers, 
$X$ which is characterized by the eigenvalue spectrum 
\begin{eqnarray}
\rho(\lambda)=\left \{
\begin{array}{ll}
(1-\alpha) \delta(\lambda)+\alpha \delta(\lambda-1), &
(0\le \alpha \le 1), \cr
\delta(\lambda-\alpha),& (\alpha > 1), 
\end{array}
\right .
\label{optdist}
\end{eqnarray}
maximizes the mutual information 
between $\vec{\weight}$ and $\vec{\out}$, 
$\forall{\alpha} \ge 0$ under the condition that 
the power of each column in $X$ is equally constrained to $\alpha$ 
\cite{Kitagawa2008}. 
For $0 \le \alpha \le 1$, this spectrum can be realized 
by composing $X$ of $p=N \alpha$ randomly 
chosen orthonormal row vectors. 
On the other hand, for $\alpha>1$, patterns of randomly constructed 
$N$ orthogonal column vectors of  dimension $N \alpha $
and length $\sqrt{\alpha}$
satisfy equation (\ref{optdist}). 
A set of row vectors of such pattern matrices
is sometimes referred to as Welch bound equality (WBE) sequences 
\cite{Welch1974}. Notice that these sequences achieve the upper-bound 
of equation (\ref{upperbound}) in the asymptotic region since 
$\left \langle \lambda^{-1} \right \rangle
=\left \langle \lambda \right \rangle^{-1}$ holds. 
This and the formal similarity between 
the channel problem and perceptron learning imply
that pattern matrices characterized by equation (\ref{optdist})
may maximize the learning performance of 
perceptrons $\forall{\alpha} \ge 0$ as well. 
Therefore, as the final example, we analyze 
the case of equation (\ref{optdist}), utilizing the 
methodology of equation (\ref{replica_entropy})
in the non-asymptotic region of $\alpha \sim O(1)$
and compare its learning performance to that of 
the i.i.d. patterns to investigate optimality.

\begin{figure}[t]
 \begin{tabular}{cc}
  \begin{minipage}{0.5\hsize}
(a)
   \begin{center}
\includegraphics[width=7cm,height=5.1cm]{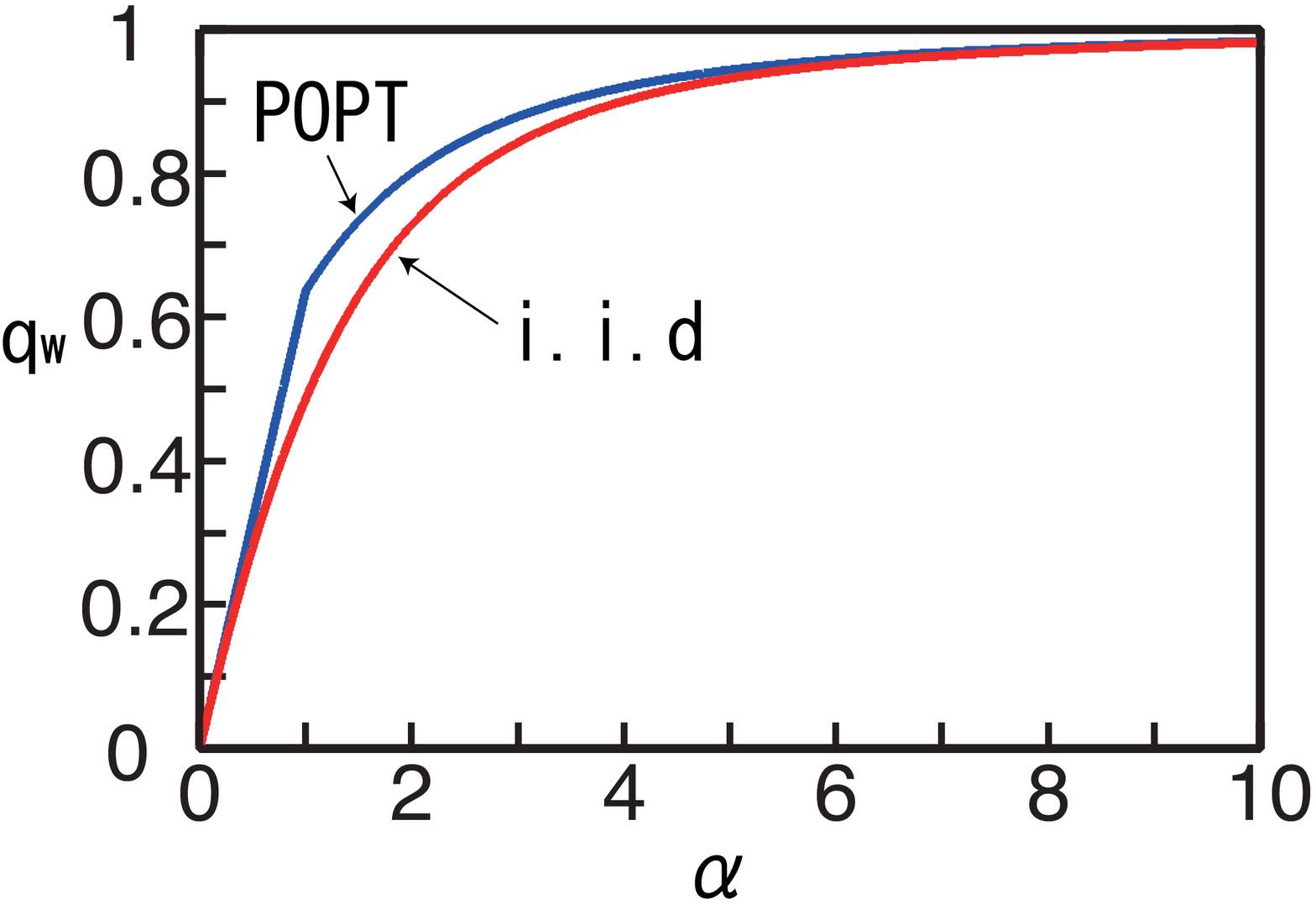} 

   \end{center}
  \end{minipage}
  \begin{minipage}{0.5\hsize}
(b)
   \begin{center}
\includegraphics[width=7cm,height=5.1cm]{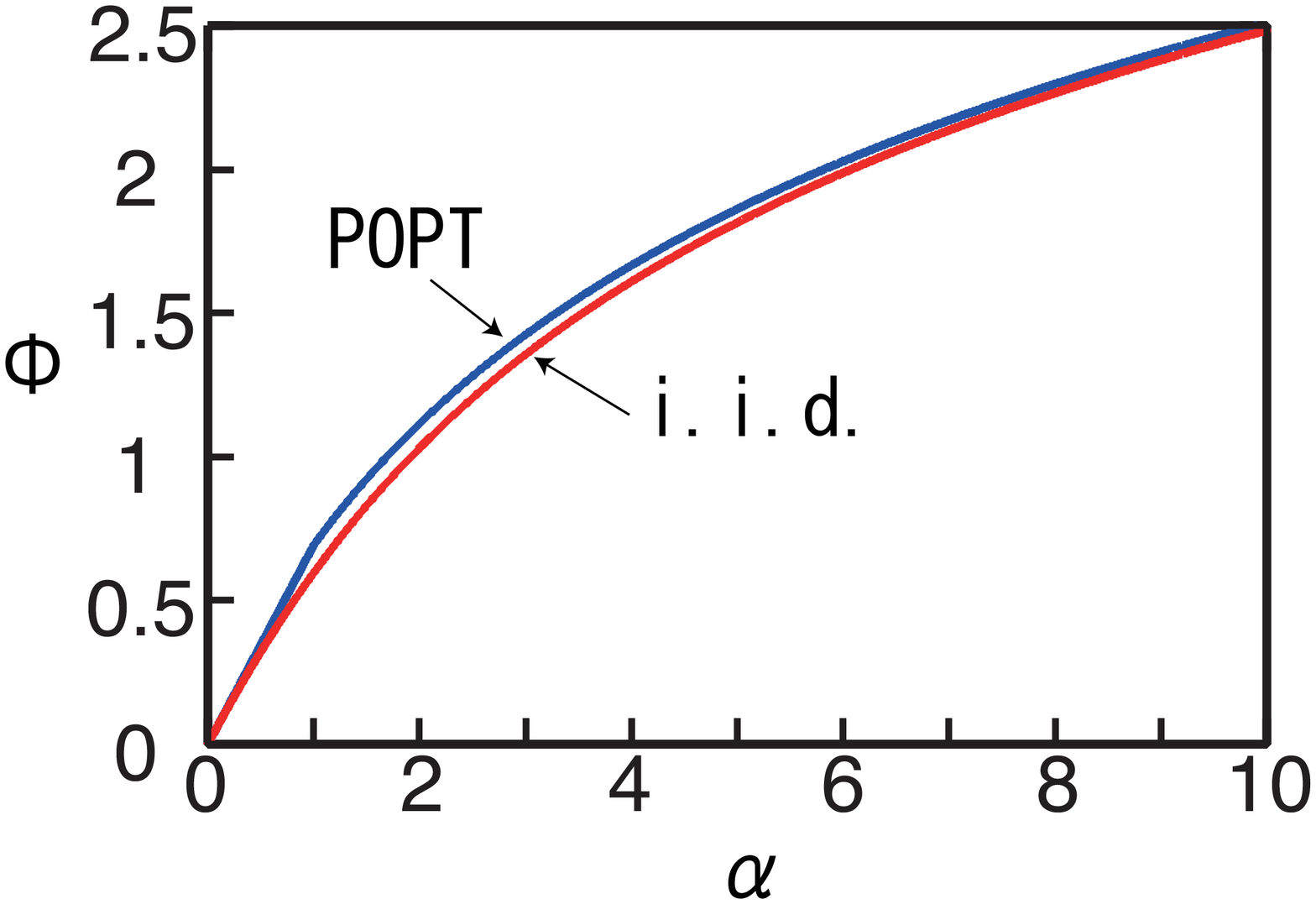}
   \end{center}
  \end{minipage}
 \end{tabular}
\caption{Performance curves for spherical weights. 
(a) $q_w$ vs. $\alpha$.\label{orthogonal_vs_iid_spherical} 
(b) $\Phi$ vs. $\alpha$. ``POPT'' and ``i.i.d.'' denote
data for the presumed optimal and i.i.d. patterns, respectively. 
Similarly for other figures. 
}
\end{figure}

For spherical weights, the saddle point problem of 
equation (\ref{replica_entropy}) can be analytically solved
for equation (\ref{optdist}) for $0 \le \alpha \le 1$ , yielding the solution
$q_\weight=2\alpha/\pi$, $q_u=2/\pi$, which in turn implies that
\begin{eqnarray}
\Phi=\alpha \log 2. 
\label{optmutual_info}
\end{eqnarray}
For $\alpha > 1$, analytical construction of the solution is
difficult and we resorted to a numerical method. 
Figures \ref{orthogonal_vs_iid_spherical} (a) and (b) 
show a comparison of the learning performance 
between the (presumed optimal) case of equation (\ref{optdist})
and the i.i.d. patterns of equation (\ref{MP}). 
For both the teacher-student overlap $q_\weight$ (figure 
\ref{orthogonal_vs_iid_spherical} (a)) and 
mutual information (free energy) $\Phi$ (figure 
\ref{orthogonal_vs_iid_spherical} (b)), 
equation (\ref{optdist}) results in
better learning performance than that of the i.i.d. patterns 
over the entire region of $\alpha \ge 0$. 

\begin{figure}[t]
 \begin{tabular}{cc}
  \begin{minipage}{0.5\hsize}
(a)
   \begin{center}
\includegraphics[width=7cm,height=5cm]{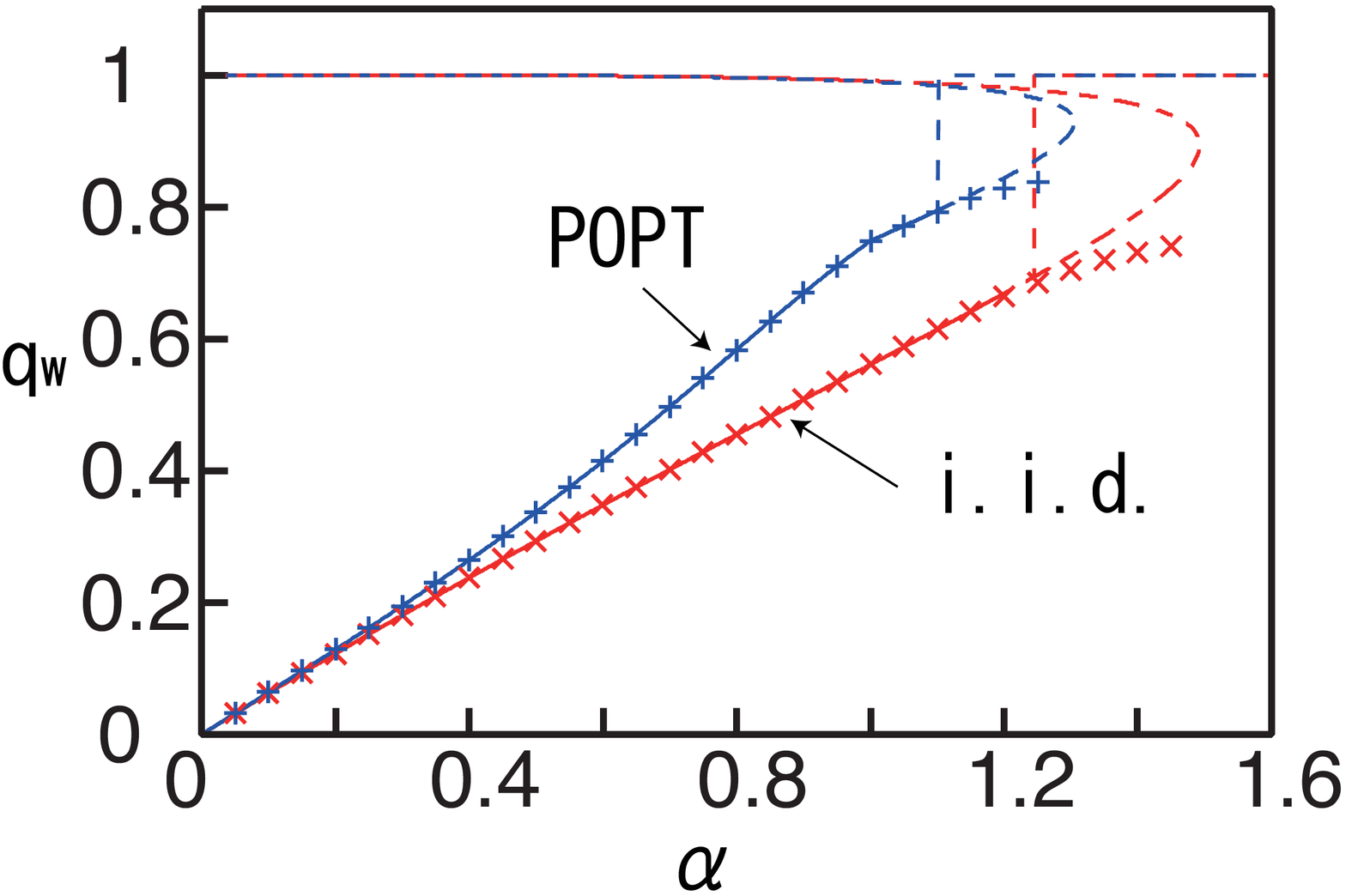} 

   \end{center}
  \end{minipage}
  \begin{minipage}{0.5\hsize}
(b)
   \begin{center}
\includegraphics[width=7cm,height=5.4cm]{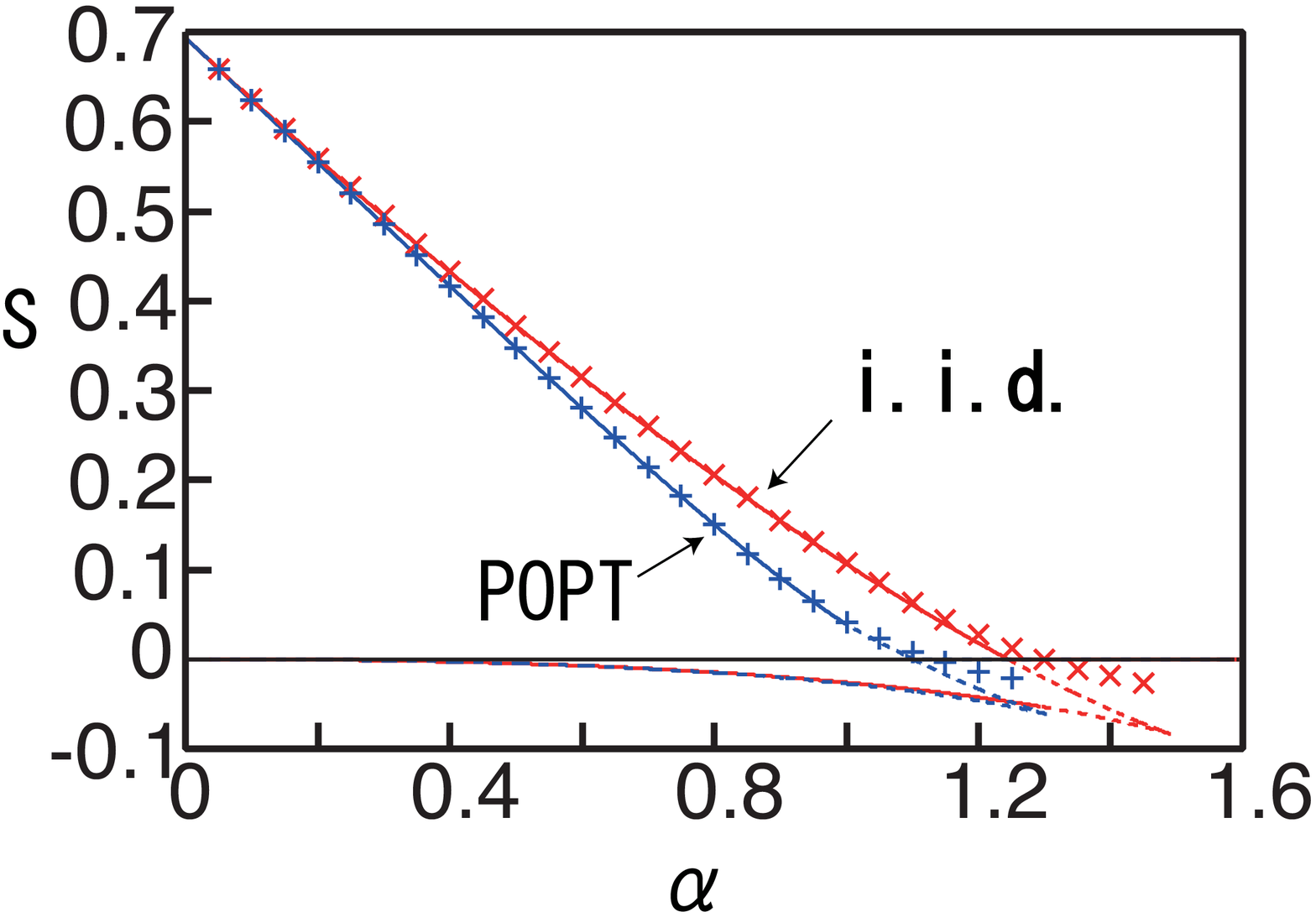}
   \end{center}
  \end{minipage}
 \end{tabular}
\caption{(a) $q_w$ vs. $\alpha$.\label{binary} (b) $S$ vs. $\alpha$.
The curves represent the theoretical prediction evaluated 
by the replica method. The markers 
are obtained from $10^4$ experiments 
for $N=100$ systems utilizing a Thouless-Anderson-Palmer 
type mean field method \cite{Shinzato2008}. 
}
\end{figure}

As another representative learning model, we examined the case 
of binary (Ising) weights, the results of which 
are shown in figures \ref{binary} (a) and (b).
For the i.i.d. patterns, it is known that simple perceptrons
with binary weights exhibit perfect learning 
at $\alpha=\alpha_c^{\rm i.i.d.} \simeq 1.245$,
completely identifying the teacher network \cite{Gyorgyi1990}. 
Such behavior is also observed for the case
of equation  (\ref{optdist}) at a certain 
critical ratio $\alpha_c^{\rm POPT}$, 
which is characterized by the vanishing entropy condition 
$S=S_0-\Phi=\log 2 -\Phi=0$. 
Figure \ref{binary} (a) yields 
$\alpha_c^{\rm POPT} \simeq 1.101$ for equation (\ref{optdist}), 
implying a better learning performance 
than that of the i.i.d. patterns. 
In terms of $q_\weight$, patterns of equation (\ref{optdist}) 
are also superior to the i.i.d. patterns (figure \ref{binary} (b)). 
In both figures \ref{binary} (a) and (b), 
numerical data for $N=100$ systems obtained from 
$10^4$ experiments based on a Thouless-Anderson-Palmer
type mean field method \cite{Shinzato2008,TAP1978} are in very close agreement
with the saddle point solutions of 
equation (\ref{replica_entropy}) (curves), 
which justifies the methodology based on equation (\ref{replica_entropy}).

\begin{figure}[htbp]
   \begin{center}
\includegraphics[width=10cm,height=8cm]{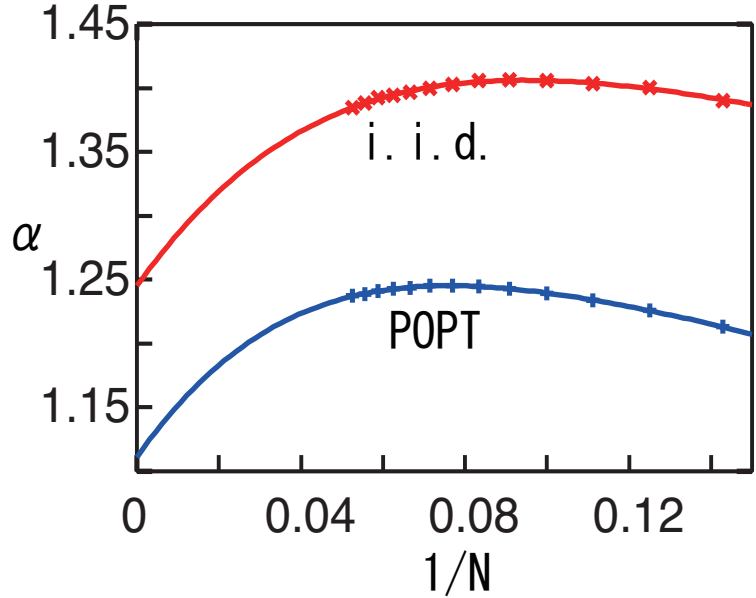} 
\caption{
Critical ratio of perfect learning 
estimated by exhaustive search experiments. 
Data (markers) are obtained by $10^6$ experiments for 
systems of $N=6,\cdots,19$. 
We fitted a fourth degree polynomial with respect to 
$1/N$, which is supported by a model selection scheme
based on the leave-one-out cross validation, 
for assessing the values as $N \to \infty$. 
Assessed values are $\lim_{N \to \infty}\alpha_{c}^{\rm POPT}(N)\simeq1.111$
and $\lim_{N \to \infty}\alpha_{c}^{\rm i.i.d.}(N)\simeq1.245$
for the presumably optimal and i.i.d. patterns, respectively. 
These are in very close agreement with the theoretical predictions 
$\alpha_{c}^{\rm POPT}\simeq1.101$ and $\alpha_{c}^{\rm i.i.d.}\simeq1.245$. 
\label{exhaustive}}
\end{center}
\end{figure}

The superiority of equation (\ref{optdist}) is also confirmed 
by experimental assessment of $\alpha_c^{\rm POPT}$. 
Figure \ref{exhaustive} shows the result of 
exhaustive search experiments for small systems 
($6 \le N \le 19$). In order to characterize
the critical ratio for finite systems, 
for each pair of $N$ and $p$, 
we estimated the probability $r(N,p)$
that at least one weight vector $\vec{\weight}$
that differs from the teacher vector $\vec{\weight}_0$
is completely compatible with a given reference data set $\xi^p$, 
utilizing $10^6$ experiments. 
For each $N$, the critical ratio is defined 
as $\alpha_c^{\rm POPT}(N)=N^{-1}\sum_{p=1}^{p_{\rm max}}r(N,p)$, 
where $p_{\rm max}$ is a sufficiently large threshold value to truncate the 
summation. We set $p_{\rm max}=4N$. 
$\alpha_c^{\rm POPT}(N)$ is expected to converge to 
$\alpha_c^{\rm POPT}$ as $N$ tends to infinity. 
In figure \ref{exhaustive}, the data plotted versus $1/N$ 
are asymmetric either side of a peak, which implies that it is
necessary to use a higher order polynomial for estimating 
$\lim_{N \to \infty} \alpha_c^{\rm POPT}(N)$
by extrapolation. 
Therefore, we fitted a fourth degree polynomial, 
which is supported by minimization of the leave-one-out 
cross validation error. The value $\lim_{N \to \infty} 
\alpha_c^{\rm POPT}(N) \simeq 1.111$ assessed by extrapolation 
agrees closely with the theoretical 
estimate $\alpha_c^{\rm POPT} \simeq 1.101$. 
This is considerably smaller than the counterpart of 
the i.i.d. patterns, $\lim_{N \to \infty} 
\alpha_c^{\rm i.i.d.}(N) \simeq 1.245$, the theoretical 
value of which is $\alpha_c^{\rm i.i.d.} \simeq 1.245$, 
indicating the superiority of equation (\ref{optdist}).

In conclusion, the above analyses for spherical and binary weights
indicate that the eigenvalue spectrum of equation (\ref{optdist}) always yields
a better learning performance than that of the i.i.d. patterns.
This lends some support to our conjecture that equation (\ref{optdist}) 
achieves optimal learning performance for perceptrons operating 
under a fixed power constraint on the pattern matrix.

\section{Summary}
In summary, we have investigated the learning performance of 
simple perceptrons extending a methodology 
for handling correlated patterns developed
in \cite{Kabashima2007,Shinzato2008} to a teacher-student scenario. 
The scheme allows us to characterize various second-order 
correlations among the input patterns by an eigenvalue 
spectrum of the cross-correlation matrix under an assumption 
that the right and left eigen-bases of the pattern 
matrix are independently generated from the Haar measure. 
Using this characterization, we have offered 
a general formula that relates the eigenvalue spectrum to 
the average free energy, which, in the current context, 
is a measure of the mutual information between 
the weight vector and output labels, given a typical pattern matrix.
The formula is used to examine cases for which column or row 
vectors in the pattern matrix are orthogonalized under a fixed 
power constraint, the learning performance of which is optimal 
for the asymptotic region and presumed to be optimal in general. 
Results from numerical experiments based on a Thouless-Anderson-Palmer 
type mean field method and exhaustive search examinations for small systems
are in agreement with theoretical predictions obtained from the formula. 

A mathematical proof of the optimality of the eigenvalue
spectrum (\ref{optdist}) and applications of the current 
scheme to various problems in learning and communication 
are promising future research directions. 
\ack

The authors thank  A. Pagnani, M. Weigt and S. Watanabe
for several exhaustive discussions and fruitful comments. 
This work was partly supported by the JSPS Global COE 
program ``Computationism as a Foundation for the Sciences'' (TS and YK) and 
Grants-in-Aid MEXT, Japan, No. 18079006 (YK).


\section*{References}
\begin{thebibliography}{99}

\bibitem{Levin1990}
  {Levin E, Tishby N and Solla S A} {1990}
  {\em Proc. IEEE} {\bf 78}, {1568}

\bibitem{Watkin1993}
  {Watkin T L H, Rau A and Biehl M}
  {1993} {\it Rev. Mod. Phys.} {\bf 65} {499}

\bibitem{Engel2001}
  {Engel A and van den Broeck C} {2001}
  {\it Statistical Mechanics of Learning} (Cambridge: Cambridge University Press)

\bibitem{Nishimori2001}
  {Nishimori H} {2001}
  {\it Statistical Physics of Spin Glasses and Information Processing - An Introduction}
  (Oxford: Oxford University Press)


\bibitem{Gardner1988}
  {Gardner E}
  {1988} {\JPA} {\bf 21} {257} 


\bibitem{Gyorgyi1990PRA}
{Gy$\ddot{\rm o}$rgyi G}
{1990} {\it Phy. Rev.} A {\bf 41}
{7097}

\bibitem{Gyorgyi1990}
  {Gy\"{o}rgyi G and Tishby N}
  {1990} 
  {\em Neural Networks and Spin Glasses} ed Theumann W K and K\"{o}berle R 
  (Singapore: World Scientific) {p 3}



\bibitem{Krauth1989}
  {Krauth W and M\'{e}zard M}
  {1989} {\em J. Physique} {\bf 50} {3056}

\bibitem{Krauth1989JPA}
  {Krauth W and Opper M}
  {1989} \JPA {\bf 22} {L519}


\bibitem{Opper1996}
  {Opper M and Kinzel W}
  {1996} 
  {\em Models of Neural Networks III} ed Domany E, van Hemmen J L and 
  Schulten K (New York: Springer-Verlag New York) p 151

\bibitem{Kabashima2003}  
  {Kabashima Y}
  {2003} \JPA {\bf 36} {11111}

\bibitem{Uda2005}
  {Uda S and Kabashima Y}
  {2005} {\em J. Phys. Soc. Jpn.} {\bf 74} {2233} 

\bibitem{Braunstein2006}
  {Braunstein A and Zecchina R}
  {2006} {\em Phys. Rev. Lett.} {\bf 96} 030201

\bibitem{Kabashima2007}
  {Kabashima Y} {2008} {\em J. Phys. Conf. Ser.} {\bf 95} {012001}

\bibitem{Shinzato2008}
{Shinzato T and Kabashima Y} 
{2008} 
{\it J. Phys. A. Math. Theor.} {\bf 41} 324013

\bibitem{Hosaka2002}
  {Hosaka T, Kabashima Y and Nishimori H}
  {2002} {\em Phys. Rev.} E {\bf 66} {066126}
\bibitem{Kinzel2002}
  {Kinzel W and Kanter I}
  {2002} {\em Proc. ICONIP'02}, Vol. 3, p 1351
\bibitem{Mimura2006}
  {Mimura K and Okada M}
  {2006} {\em Phys. Rev.} E {\bf 74} {026108}

\bibitem{Kabashima2008}
  {Kabashima Y} {2008}
  {An integral formula for large random rectangular matrices and
    its application to analysis of linear vector channels}
{\em Proc. The 1st Workshop on Physics-Inspired Paradigms in Wireless Communications and Networks} (Berlin, Germany) ({\em Preprint} {arXiv:0802.1372})


\bibitem{Nishimori1981}
{Nishimori H}
{1981}
{\it Prog. Theor. Phys.} {\bf 66}
{1169}

\bibitem{Nishimori1993}
{Nishimori H}
{1993}
{\it J. Phys. Soc. Jpn.}
{\bf 62}
{2973}

\bibitem{Nishimori2001b}
{Nishimori H and Sherrington D}
{2001}
{\em Disordered and Complex Systems} ed Sollich P, Coolen A C C, Hughston L P and Streater R F (New York: AIP) {p 67}

\bibitem{Baum1990}
{Baum E B and Haussler D}
{1990}
{\em Neural Comput.} {\bf 1}
{151}

\bibitem{Amari1992}
{Amari S, Fujita N and Shinomoto S}
{1992}
{\em Neural Comput.} {\bf 4}
{605}

\bibitem{Murata1994}
{Murata N, Yoshizawa S and Amari S }
{1994}
{\em IEEE Trans. Neural Net.}
{865}

\bibitem{Watanabe2001}
{Watanabe S}
{2001}
{\em Neural Comput. } {\bf 13}
{899}

\bibitem{Fukumizu1994}
{Fukumizu K}
{1996}
{\em Advances in Neural Information Processing Systems} {\bf 8}
{312}

\bibitem{Sollich1994}
{Sollich P}
{1994}
{\em Phys. Rev.} {E} {\bf 49}
{4637}

\bibitem{Seeger2008}
{Seeger M W}
{2008}
{\em Journal of Machine Learning Research}
{\bf 9}
{759} 

\bibitem{Kitagawa2008}
{Kitagawa K and Tanaka T}
{2008}
{\em Proceedings of the 2008 IEEE International Symposium on Information Theory}
(Tronto, Canada) p 
{1373}

\bibitem{Welch1974}
{Welch L R}
{1974}
{\em IEEE Trans. Info. Theory} {\bf IT-20}
{397}

\bibitem{TAP1978}
  {Thouless D J, Anderson P W and Palmer R G}
  {1977} {\em Phil. Mag.} {\bf 35} {593}








\end{thebibliography}
\end{document}